\documentclass{Odyssey2026}

\interspeechcameraready

\title{From Self-Supervised Speech Models to Mixture-of-Experts for Robust Anti-Spoofing}

\author[affiliation={1,2}]{Hugo}{Daumain}
\author[affiliation={1}]{Driss}{Matrouf}
\author[affiliation={2}]{Khaled}{Khelif}
\author[affiliation={1}]{Mickael}{Rouvier}

\affiliation{LIA}{Université d'Avignon}{France}
\affiliation{}{Airbus Defence \& Space}{France}
\email{hugo.daumain@alumni.univ-avignon.fr}


\usepackage{comment}
\usepackage{multirow}
\usepackage{soul}
\usepackage{float}
\usepackage{tikz}
\usetikzlibrary{positioning,,arrows.meta,fit,calc,decorations.pathreplacing}

\usepackage{booktabs}
\usepackage[table]{xcolor}
\usepackage{colortbl}
\usepackage{siunitx}

\usepackage{dblfloatfix}
\usepackage{numprint}
\usepackage{subcaption}

\usepackage{pgfplots}
\pgfplotsset{compat=1.18}
\usepackage{xcolor}
\usepgfplotslibrary{groupplots}
 
\definecolor{expA}{HTML}{4C78A8}
\definecolor{expB}{HTML}{E15759}
\definecolor{expC}{HTML}{72B7B2}
\definecolor{expD}{HTML}{F28E2B}
\usetikzlibrary{backgrounds}
\pgfdeclarelayer{bg}
\pgfsetlayers{bg,main}

\pgfplotsset{
  layerstyle/.style={
    ybar=0.1pt,
    bar width=4.5pt,
    ymin=0, ymax=0.88,
    xmin=-5, xmax=65,
    ytick={0,0.25,0.50,0.75},
    yticklabel style={font=\tiny,
      /pgf/number format/fixed,
      /pgf/number format/precision=2},
    ylabel style={font=\tiny, yshift=1pt},
    xtick={0,10,20,30,40,50,60},
    xticklabels={AudioGen, FlashSpeech, NaturalSpeech3, OpenAI, PromptTTS2, VALLE, xTTS},
    xticklabel style={font=\fontsize{4}{6}\selectfont, rotate=35, anchor=east},
    width=7.2cm,
    height=2.0cm,
    scale only axis,
    axis x line=bottom,
    axis y line=left,
    axis line style={gray!60, line width=0.4pt},
    tick style={gray!60, line width=0.3pt},
    ymajorgrids=false,
    xmajorgrids=false,
    every axis plot/.append style={draw=white, line width=0.3pt},
    clip=false,
  },
}
\npdecimalsign{.}
\newcommand{\eer}[1]{\nprounddigits{2}\numprint{#1}}
\newcommand{\js}[1]{\nprounddigits{3}\numprint{#1}}

\usepackage{xfp}

\newcommand{\params}[1]{%
  \nprounddigits{0}\numprint{\fpeval{#1/1000000}}M%
}

\begin{document}

\maketitle

\begin{abstract}

Recent advances in speech generation have significantly improved the naturalness of synthetic speech, making spoofing detection increasingly challenging. A key limitation of current anti-spoofing systems is their limited robustness to unseen synthesis methods. In this work, we transform a self-supervised speech representation model into a Mixture-of-Experts (MoE) architecture to improve generalization. Feed-forward blocks in selected encoder layers are replaced by multiple expert networks controlled by a layer-wise gating mechanism, allowing experts to capture complementary acoustic patterns while preserving the representations learned during self-supervised pretraining. We further analyze the architectural choices affecting the performance of this MoE conversion and investigate the activation behavior of the experts. The proposed approach is evaluated on 14 spoofing datasets and reduces the macro EER from 5.46\% to 4.81\%, corresponding to 11.9\% relative improvement over the baseline.


\end{abstract}

\section{Introduction}


In recent years, speech spoofing technologies have made significant progress. Modern voice conversion systems and text-to-speech synthesizers now rely on neural audio codec~\cite{wang2023neural}, flow-matching approaches~\cite{mehta2024matcha}, and diffusion-based architectures~\cite{popov2021grad}, producing speech of high perceptual quality. As a result, distinguishing bonafide from spoofed speech (synthetic or manipulated speech) has become increasingly challenging. This technological progress also raises critical societal concerns as highly realistic speech synthesis can be exploited for impersonation, large-scale misinformation, and public opinion manipulation~\cite{vaccari2020deepfakes, yi2023audio}. 
Developing reliable speech anti-spoofing systems has therefore become a research priority.

Classical countermeasure approaches rely on supervised deep neural networks (DNNs) trained to detect artifacts left by speech generation systems. These artifacts manifest as residual traces in the temporal and spectral structure of the signal, often imperceptible to human listeners.
Early end-to-end approaches focused on convolutional architectures operating directly on raw waveforms, such as SincNet~\cite{ravanelli2018interpretable} and RawNet2~\cite{tak2021end}, which learn discriminative temporal representations for spoofing detection.
More recent supervised architectures introduce explicit modeling of spectro-temporal dependencies to better capture these artifacts. In particular, graph-based approaches such as RAWGAT-ST~\cite{tak2021end2} and AASIST~\cite{jung2022aasist} leverage graph attention mechanisms to capture structured relationships across time and frequency components of the acoustic signal.

However, a fundamental limitation persists: spoofing artifacts are often synthesizer-dependent, and models trained on a given set of generation techniques may fail to generalize to unseen attacks. In practical scenarios, where attack methods continuously evolve, robustness to distribution shifts becomes the central challenge.

Recently, Self-Supervised Learning (SSL) models such as Wav2vec2~\cite{baevski2020wav2vec}, WavLM~\cite{chen2022wavlm} or HuBERT~\cite{hsu2021hubert} have emerged as powerful backbone architectures for speech processing tasks. Pre-trained on large-scale unlabeled corpora, these models capture rich acoustic and linguistic representations, yielding highly transferable representations. Their strong generalization capacity has made them particularly attractive for anti-spoofing systems~\cite{li2025measuring}.

Beyond representation learning, recent work has explored architectural strategies to further enhance model capacity, among which Mixture-of-Experts (MoE) has gained increasing attention. 
MoE architectures scale model capacity through conditional expert activation~\cite{huang2024toward}, making them well suited to heterogeneous data such as spoofing artifacts, which vary across synthesizers, languages, and acoustic conditions~\cite{mu2025comprehensive}.
Furthermore, recent studies have shown that dense pretrained models can be converted into MoE architectures through weight reuse, preserving pretrained knowledge while expanding model capacity~\cite{komatsuzaki2022sparse, fu2025ume}.

In speech anti-spoofing, MoE approaches have recently been explored at various architectural levels. 
Some introduce MoE at the model level, treating experts as complete detectors, with either identical~\cite{negroni2025leveraging} or diverse~\cite{negroni2025attention} lightweight architectures. 
Within SSL-based systems, MoE has been applied in two main ways: as a mechanism to aggregate multi-layer representations extracted from the backbone~\cite{wang2025mixture, hao2025wav2df}, or as a set of low-rank (LoRA) modules~\cite{hu2022lora} integrated directly into the SSL architecture~\cite{laakkonen2025mixture, pan2025molex, chen2025adaptive}.
In the latter setting, each expert is implemented as a low-rank weight correction applied to selected linear layers of a frozen pretrained model. A routing mechanism dynamically selects or combines these corrections depending on the input, enabling input-dependent modulation of the SSL representations. Since these adaptations are parameterized in low rank and the backbone remains frozen, only a small number of additional parameters need to be learned, resulting in efficient and stable task adaptation.
However, by constraining each expert to a low-rank correction, these strategies can limit the degree to which expert specialization can reshape internal representations.



In this work, we explore the conversion of a self-supervised speech model into a full Mixture-of-Experts (MoE) architecture to improve generalization in speech anti-spoofing. Feed-forward blocks in selected encoder layers are replaced by multiple expert networks controlled by a layer-wise gating mechanism, encouraging experts to capture complementary spoofing-related patterns while preserving the representations learned during self-supervised pretraining.

The main contributions of this paper are:
\begin{itemize}
    \item \textbf{A full MoE conversion paradigm for speech anti-spoofing:} to the best of our knowledge, this is the first work to investigate the conversion of a pretrained SSL speech model into a full Mixture-of-Experts architecture for this task, rather than relying on LoRA-based expert adaptation.
    \item \textbf{An extensive architectural study:} we analyze the impact of several key design choices, including expert placement, number of experts, and pooling strategy for the gating network.
    \item \textbf{An analysis of expert activation behavior:} We examine whether the best-performing MoE configuration shows signs of expert specialization, particularly with regard to its inter-synthesizer behavior.
\end{itemize}

Evaluated on 14 spoofing datasets, the best MoE approach based on WavLM-Large reduces the macro EER from 5.46\% to 4.81\%, corresponding to 11.9\% relative improvement over the baseline.

This article is organized as follows. 
We present a state-of-the-art SSL-based anti-spoofing architecture in Section~\ref{sec:baseline} and the proposed Mixture-of-Experts (MoE) conversion in Section~\ref{sec:moe}.
After describing the experimental setup in Section~\ref{sec:exp_setup}, we present the results of various experiments on architectural choices in Section~\ref{sec:results} and analyze potential expert specialization in Section~\ref{sec:analysis}.
Finally, Section~\ref{sec:conclusion} concludes this work.

\section{State-of-the-Art Approach}
\label{sec:baseline}
The current state-of-the-art approaches to anti-spoofing are based on a neural architecture composed of a self-supervised learning model, which extracts rich feature representations from the input speech signal (Section~\ref{sec:ssl}), and a classifier, which discriminates between bonafide and spoofed speech signals (Section~\ref{sec:mhfa}).


\subsection{Self-Supervised Learning model} 
\label{sec:ssl}

Self-Supervised Learning (SSL) models are considered as powerful backbone architectures for speech anti-spoofing. Unlike conventional supervised approaches that learn task-specific representations from limited labeled spoofing datasets, SSL models are pre-trained on large-scale unlabeled speech corpora. This pre-training enables the extraction of rich, hierarchical representations that encode fine-grained acoustic structure as well as higher-level phonetic and prosodic information.

Most of these SSL models follow a similar two-stage design (Wav2vec2, HuBERT, WavLM,...) composed of a convolutional feature extractor followed by a transformer encoder. Let $x \in \mathbb{R}^{T_{raw}}$ be a raw waveform, the feature extractor implemented as a stack of strided 1D convolutions produces a latent sequence $H_0 \in \mathbb{R}^{T \times F_{cnn}}, \quad T < T_{raw}$.
This sequence is then processed by $L$ transformer encoder layers. Each layer $l \in \{1, \dots, L\}$ is based on a multi-head self-attention (MHA) mechanism and a feed-forward network coupled with residual connections and layer normalization (Figure~\ref{fig:wavlm_moe_comparison}). The output representation $H_l \in \mathbb{R}^{T \times F}$ of each layer $l$ captures signal-level information in the lower layers, while higher layers encode progressively richer and more abstract representations.

\subsection{Classification}
\label{sec:mhfa}

\begin{figure}[t]
\centering
\begin{tikzpicture}[
    font=\footnotesize,
    scale=0.75, transform shape,
    >=stealth,
    node distance=1.2mm,
    frame/.style={draw, dashed, thick, rounded corners=2pt, inner sep=2.0mm},
    layer/.style={rectangle, draw, very thick, rounded corners=2pt,
                  minimum width=2.05cm, minimum height=5.2mm, align=center, fill=blue!18},
    cnn/.style={rectangle, draw, very thick, rounded corners=2pt,
                minimum width=2.05cm, minimum height=5.2mm, align=center, fill=red!30},
    dots/.style={draw=none, minimum width=2.05cm, minimum height=3.0mm, align=center},
    mhfaBox/.style={rectangle, draw, thick, rounded corners=2pt,
                    minimum width=1.05cm, minimum height=5.2mm, align=center, fill=gray!10},
    mhfaBig/.style={rectangle, draw, thick, rounded corners=2pt,
                    minimum width=1.75cm, minimum height=7.2mm, align=center, fill=gray!15},
    mhfaBack/.style={rectangle, draw, thick, rounded corners=2pt,
                 minimum width=1.75cm, minimum height=7.2mm, align=center, fill=gray!8},
    sumNode/.style={circle, draw, thick, inner sep=0pt, minimum size=4.2mm, fill=white},
    coeffV/.style={circle, draw, thick, inner sep=0pt, minimum size=4.0mm, fill=yellow!20},
    coeffK/.style={circle, draw, thick, inner sep=0pt, minimum size=4.0mm, fill=violet!18},
    qbox/.style={rectangle, draw, thick, rounded corners=2pt,
                 minimum width=0.72cm, minimum height=4.2mm, align=center, fill=orange!18},
    pred/.style={rectangle, draw, thick, rounded corners=2pt,
                 minimum width=1.55cm, minimum height=6.2mm, align=center},
    arr/.style={->, line width=0.7pt},
    smallarr/.style={->, line width=0.4pt},
    varr/.style={->, line width=0.9pt, draw=yellow!65!black},
    karr/.style={->, line width=0.9pt, draw=violet!75!black},
    qarr/.style={->, line width=0.85pt, draw=orange!85!black},
]

\node[cnn] (cnn0) at (0,0) {CNN Encoder};
\node[layer] (l1) at (0,1.20) {Transformer 1};
\node[dots] (d1) at (0,2.40) {$\vdots$};
\node[layer] (l13) at (0,3.45) {Transformer L};
\node[frame, fit=(cnn0)(l13), label={[font=\bfseries]above:SSL model}] {};
\draw[smallarr] (cnn0.north) -- (l1.south);
\draw[smallarr] (l1.north) -- ($(d1.south)+(0,-0.10)$);
\draw[smallarr] ($(d1.north)+(0,-0.10)$) -- (l13.south);

\node[sumNode] (sumK) at ($(l13.east)+(3.00cm,0.20cm)$) {$+$};
\node[mhfaBox, right=8mm of sumK] (linK) {Linear};
\node[qbox, above left=4.5mm and 2mm of linK] (q1) {$Q_1$};
\node[qbox, right=0.9mm of q1] (q2) {$Q_2$};
\node[qbox, right=0.9mm of q2] (q3) {$\cdots$};
\node[qbox, right=0.9mm of q3] (qh) {$Q_H$};
\node[draw=orange!85!black, rounded corners=2pt, thick,
      fit=(q1)(q2)(q3)(qh), inner sep=1.2pt] (qblock) {};
\node[mhfaBack, right=15mm of linK, xshift=2.2mm, yshift=2.2mm] (attnback1) {};
\node[mhfaBack, right=15mm of linK, xshift=1.1mm, yshift=1.1mm] (attnback2) {};
\node[mhfaBig, right=15mm of linK] (attn) {Head-wise\\attention};
\node[sumNode, below=9.5mm of sumK] (sumV) {$+$};
\node[mhfaBox, right=8mm of sumV] (linV) {Linear};
\node[mhfaBack, below=6.5mm of attn, xshift=2.2mm, yshift=2.2mm] (poolback1) {};
\node[mhfaBack, below=6.5mm of attn, xshift=1.1mm, yshift=1.1mm] (poolback2) {};
\node[mhfaBig, below=6.5mm of attn] (pool) {Attentive\\pooling};
\node[mhfaBox, below=2.0mm of pool] (concat) {Concat};
\node[mhfaBox, below=2.0mm of concat] (linOut) {Linear};
\node[pred, below=2.3mm of linOut] (pred) {\bfseries Prediction\\[-0.4mm]\scriptsize spoof/bonafide};

\draw[decorate,decoration={brace,amplitude=4pt}]
  ($(attnback1.south east)+(0.75mm,0.75mm)$) --
  ($(attn.south east)+(-0.75mm,-0.75mm)$)
  node[midway,xshift=3mm, yshift=-2.2mm] {$H$};
\draw[decorate,decoration={brace,amplitude=4pt}]
  ($(poolback1.south east)+(0.75mm,0.0mm)$) --
  ($(pool.south east)+(-0.75mm,-0.75mm)$)
  node[midway,xshift=3mm, yshift=-2.2mm] {$H$};

\node[frame, fit=(sumK)(linK)(qblock)(attn)(sumV)(linV)(pool)(concat)(linOut)(pred)(q1),
inner xsep=10pt, xshift=1.5mm,
label={[font=\bfseries]above:Multi-Head Factorized Attention}] {};

\node[coeffK] (wk0)  at ($(cnn0.east)+(1.25cm, 0.32cm)$) {$w_0^K$};
\node[coeffK] (wk1)  at ($(l1.east)+(1.25cm,  0.32cm)$)  {$w_1^K$};
\node[coeffK] (wk13) at ($(l13.east)+(1.25cm, 0.32cm)$)  {$w_L^{K}$};
\node[coeffV] (wv0)  at ($(cnn0.east)+(1.25cm,-0.32cm)$) {$w_0^V$};
\node[coeffV] (wv1)  at ($(l1.east)+(1.25cm, -0.32cm)$)  {$w_1^V$};
\node[coeffV] (wv13) at ($(l13.east)+(1.25cm,-0.32cm)$)  {$w_L^{V}$};

\draw[karr] (cnn0.east) -- (wk0.west);
\draw[karr] let \p1=(wk0.east), \p2=(sumK.west), \p3=(wk13.east) in
    (wk0.east) -- (\x3+2mm,\y1) -- (\x3+2mm,\y2) -- (sumK.west);
\draw[karr] (l1.east) -- (wk1.west);
\draw[karr] let \p1=(wk1.east), \p2=(sumK.west), \p3=(wk13.east) in
    (wk1.east) -- (\x3+2mm,\y1) -- (\x3+2mm,\y2) -- (sumK.west);
\draw[karr] (l13.east) -- (wk13.west);
\draw[karr] let \p1=(wk13.east), \p2=(sumK.west) in
    (wk13.east) -- (\x1+2mm,\y1) -- (\x1+2mm,\y2) -- (sumK.west);
\draw[karr] (sumK.east) -- (linK.west);

\draw[varr] (cnn0.east) -- (wv0.west);
\draw[varr] let \p1=(wv0.east), \p2=(sumV.west), \p3=(wv13.east) in
    (wv0.east) -- (\x3+7mm,\y1) -- (\x3+7mm,\y2) -- (sumV.west);
\draw[varr] (l1.east) -- (wv1.west);
\draw[varr] let \p1=(wv1.east), \p2=(sumV.west), \p3=(wv13.east) in
    (wv1.east) -- (\x3+7mm,\y1) -- (\x3+7mm,\y2) -- (sumV.west);
\draw[varr] (l13.east) -- (wv13.west);
\draw[varr] let \p1=(wv13.east), \p2=(sumV.west) in
    (wv13.east) -- (\x1+7mm,\y1) -- (\x1+7mm,\y2) -- (sumV.west);
\draw[varr] (sumV.east) -- (linV.west);

\draw[qarr] (qblock.east) -| node[right, font=\scriptsize, black] {$Q_H$} (attn.north);
\draw[arr]  (attn.south) -- node[right, font=\scriptsize, yshift=1mm] {$A_H$} (pool.north);
\draw[varr] (linV.east)  -- node[above, font=\scriptsize, black] {$V$} (pool.west);
\draw[karr] (linK.east)  -- node[above, font=\scriptsize, black] {$K$} (attn.west);

\draw[arr] (pool.south)   -- (concat.north);
\draw[arr] (concat.south) -- (linOut.north);
\draw[arr] (linOut.south) -- (pred.north);

\node[font=\scriptsize, anchor=west, text=violet!75!black] at ($(sumV.south west)+(-0.10,-1.20)$) {\rule{3mm}{0.7mm} Key flow};
\node[font=\scriptsize, anchor=west, text=yellow!65!black] at ($(sumV.south west)+(-0.10,-1.75)$) {\rule{3mm}{0.7mm} Value flow};
\node[font=\scriptsize, anchor=west, text=orange!85!black] at ($(sumV.south west)+(-0.10,-2.30)$) {\rule{3mm}{0.7mm} Learned queries};
\end{tikzpicture}
\caption{Representations extracted from the SSL feature extractor and selected Transformer layers are routed to the MHFA module. MHFA attentively aggregates multi-layer representations to produce an utterance-level embedding used for the final prediction.}
\label{fig:wavlm_mhfa_detailed}
\end{figure}
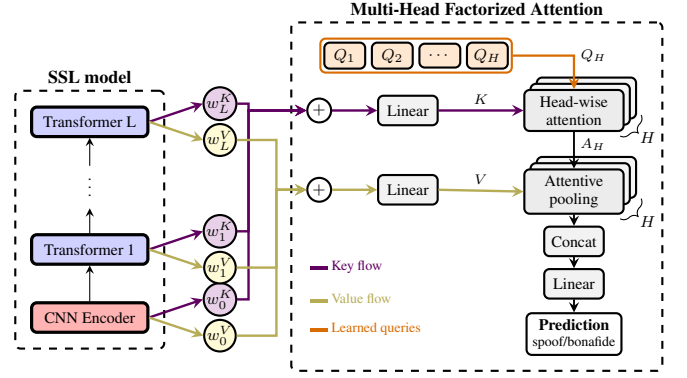

Following the SSL backbone, a Multi-Head Factorized Attention (MHFA) module~\cite{peng2023attention} is used as the classification head, as illustrated in Figure~\ref{fig:wavlm_mhfa_detailed}. Unlike approaches relying solely on the final Transformer layer, MHFA leverages representations from multiple SSL layers.
MHFA applies an attention mechanism with learnable query vectors.

Let $H_l \in \mathbb{R}^{T \times F}$ denote the output of layer $l$. The aggregated keys and values are computed as

\begin{equation}
K = \sum_{l=0}^{L} w_l^{k} H_l S_k,
\qquad
V = \sum_{l=0}^{L} w_l^{v} H_l S_v,
\end{equation}

\noindent where $S_k, S_v \in \mathbb{R}^{F \times D}$ are projection matrices, and $w_l^{k}$, $w_l^{v}$ are layer-wise scalar weights.

\noindent The attention weights and head-wise outputs are computed as

\begin{equation}
A_h = \mathrm{softmax}(KQ_h),
\qquad
Z_h = A_h^\top V,
\end{equation}

\noindent where $Q_h$ denotes a learnable query vector associated with head $h$. The head-wise outputs $Z_h$ are then pooled and concatenated to produce the final utterance-level embedding. The utterance-level embedding is finally passed through a linear layer with a sigmoid activation to produce the bonafide/spoof probability.

\section{Mixture-of-Experts}
\label{sec:moe}

Based on the classical approach described above, this section presents the proposed Mixture-of-Experts (MoE) architecture applied to selected layers of the SSL backbone. We first describe the MoE transformation of the Transformer layers (Section~\ref{sec:ssl_mixture_of_experts_layers}), then detail the gating network used for expert routing (Section~\ref{sec:gating_network}), and finally introduce the auxiliary loss used to prevent expert collapse (Section~\ref{sec:auxiliary_loss}).



\subsection{SSL Mixture-of-Experts layers}
\label{sec:ssl_mixture_of_experts_layers}

\begin{figure}[H]
\centering
\begin{subfigure}[t]{0.42\linewidth}
\centering
\begin{tikzpicture}[scale=0.55, transform shape, font=\normalsize,
arrow/.style={-{Latex[length=1.5mm,width=1.2mm]}, line width=0.6pt},
block/.style={draw, rounded corners=2pt, minimum width=3.8cm, minimum height=1cm, align=center, line width=0.9pt},
frame/.style={draw, dashed, rounded corners=2pt, inner sep=1.5mm, line width=0.9pt}
]
\node[block] (mhsa) {Multi-Head Self-Attention};
\node[block, below=4mm of mhsa] (addnorm1) {Add \& Norm};
\node[block, below=4mm of addnorm1] (ffn) {FFN};
\node[block, below=4mm of ffn] (addnorm2) {Add \& Norm};
\draw[arrow] ($(mhsa.north)+(0,8mm)$) -- (mhsa.north);
\draw[arrow] (mhsa) -- (addnorm1);
\draw[arrow] (addnorm1) -- (ffn);
\draw[arrow] (ffn) -- (addnorm2);
\draw[arrow] (addnorm2.south) -- ++(0,-7mm);
\node[frame, fit=(mhsa)(addnorm2)] {};
\end{tikzpicture}
\caption{Standard Transformer layer}
\end{subfigure}
\hspace{0.01\linewidth}
\begin{subfigure}[t]{0.54\linewidth}
\centering
\begin{tikzpicture}[scale=0.50, transform shape, font=\normalsize,
arrow/.style={-{Latex[length=1.5mm,width=1.2mm]}, line width=0.6pt},
block/.style={draw, rounded corners=2pt, minimum width=3.8cm, minimum height=1cm, align=center, line width=0.9pt},
block_expert/.style={draw, rounded corners=2pt, minimum width=1.4cm, minimum height=0.7cm, align=center, line width=0.9pt},
gate/.style={draw, rounded corners=2pt, minimum width=2.2cm, minimum height=1.0cm, align=center, line width=0.9pt},
mult/.style={circle, draw, fill=white, inner sep=0.6pt, minimum size=4mm, line width=0.7pt},
weight/.style={-{Latex[length=1mm,width=0.9mm]}, line width=0.7pt, dashed},
frame/.style={draw, dashed, rounded corners=2pt, inner sep=1.2mm, line width=0.9pt}
]
\node[block] (mhsa) {Multi-Head Self-Attention};
\node[block, below=4mm of mhsa] (addnorm1) {Add \& Norm};
\node[block_expert, below left=7mm and 2mm of addnorm1.south] (exp2) {Expert 2};
\node[block_expert, below=7mm of addnorm1, left=2mm of exp2] (exp1) {Expert 1};
\node[block_expert, below=7mm of addnorm1, right=8mm of exp2] (expn) {Expert N};
\node[below=7mm of addnorm1, right=2mm of exp2] (txt) {...};
\node[block, below=28mm of addnorm1] (addnorm2) {Add \& Norm};
\node[gate, below right=5mm and 4mm of addnorm1.east] (gating) {Gating};
\node[frame, fit=(exp1)(mhsa)(addnorm2)(gating)] {};
\node[mult, below=6mm of exp1] (m1) {$\times$};
\node[mult, below=6mm of exp2] (m2) {$\times$};
\node[mult, below=6mm of expn] (mN) {$\times$};
\draw[arrow] ($(mhsa.north)+(0,8mm)$) -- (mhsa.north);
\draw[arrow] (mhsa.south) -- (addnorm1);
\draw[arrow] (addnorm1) -- (exp1);
\draw[arrow] (addnorm1) -- (exp2);
\draw[arrow] (addnorm1) -- (expn);
\draw[arrow] (exp1) -- (m1);
\draw[arrow] (exp2) -- (m2);
\draw[arrow] (expn) -- (mN);
\draw[arrow] (m1) -- (addnorm2);
\draw[arrow] (m2) -- (addnorm2);
\draw[arrow] (mN) -- (addnorm2);
\draw[arrow] (addnorm2.south) -- ++(0,-7mm);
\draw[arrow] (addnorm1.east) -- (gating.north);
\begin{pgfonlayer}{bg}
\draw[weight] (gating.south) .. controls +(0,-4mm) and +(10mm,0) .. (m1.east);
\draw[weight] (gating.south) .. controls +(0,-8mm) and +(10mm,0) .. (m2.east);
\draw[weight] (gating.south) .. controls +(0,-12mm) and +(10mm,0) .. (mN.east);
\end{pgfonlayer}
\end{tikzpicture}
\caption{Mixture-of-Experts Transformer layer}
\end{subfigure}
\caption{Comparison between (a) a standard SSL model Transformer layer and (b) a Mixture-of-Experts Transformer layer. In (b), the feed-forward network (FFN) is replaced by $N$ parallel FFNs, each considered as an expert; a gating module computes a routing distribution and selects the Top-$K$ experts, whose outputs are combined through a weighted sum.}
\label{fig:wavlm_moe_comparison}
\end{figure}
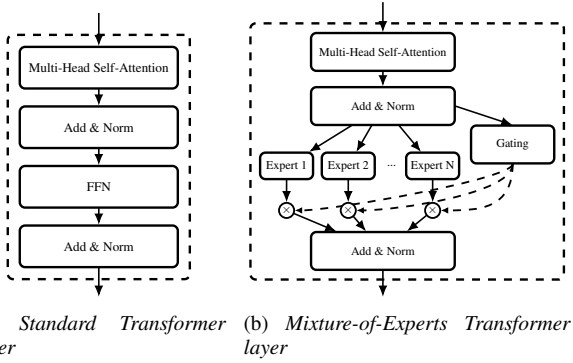

We convert a subset $\{ \mathcal{L}_i \mid i \in \mathcal{S} \subseteq \{1, \dots, L\} \}$ of SSL transformer layers into Mixture-of-Experts (MoE) layers, where each selected layer $\mathcal{L}_i$ is replaced by its MoE counterpart.
As depicted in Figure \ref{fig:wavlm_moe_comparison}, a MoE layer is obtained by replacing the original dense feed-forward module by $E$ new dense feed-forward modules considered as experts. 
Each one of these $E$ experts is initialized with the weights of the original dense feed-forward module, to prevent forgetting knowledge from the pretraining stage. 
Expert selection is performed by a gating network (described below), which computes a routing distribution over the
$E$ experts of the post-MHA representations and activates only the highest top-$k$ scoring experts for the entire utterance. This gating network is layer-dependent; each MoE layer possesses its own gating network. 

\subsection{Gating network}
\label{sec:gating_network}

The gating network assigns routing probabilities to the $E$ experts and selects the $k$ most relevant ones according to a top-$k$ routing strategy. 
To compute these routing probabilities, the frame-level representations produced by the transformer layer must first be aggregated into an utterance-level embedding.

Let $[\mathbf{h}_1, \dots, \mathbf{h}_T] \in \mathbb{R}^{T \times F}$ denote the sequence of frame-level representations after the MHA module at layer $l$. 
An utterance-level representation $\mathbf{g}$ is obtained by applying a pooling operation along the temporal dimension. We investigate several pooling strategies such as mean, max, statistical and attentive pooling. 





Given the pooled representation $\mathbf{g}$, the gating network produces routing probabilities for each expert:
\begin{equation}
\mathbf{p} = \operatorname{Softmax}(\mathbf{W}_g \mathbf{g}),
\end{equation}
where $\mathbf{p} \in \mathbb{R}^{E}$ denotes the probability of selecting each expert.
\noindent Following the top-$k$ routing strategy, only the $k$ experts with the highest probabilities are activated:
\begin{equation}
\mathcal{K} = \operatorname{TopK}(\mathbf{p}, k).
\end{equation}

\noindent  The corresponding probabilities are then re-normalized as
\begin{equation}
\tilde{p}_i = \frac{p_i}{\sum_{j \in \mathcal{K}} p_j}.
\end{equation}
and used as the combination weights of the selected experts.
\noindent The non-differentiable top-$k$ selection does not prevent training as gradients still flow through the renormalized softmax weights back to the gating module.







\subsection{Auxiliary loss}
\label{sec:auxiliary_loss}

To encourage a balanced utilization of experts and prevent expert collapse, we employ an auxiliary load-balancing loss following~\cite{fedus2022switch}. 

Let $E$ denote the number of experts. 
We define $p_i$ as the mean routing probability assigned to expert $i$, 
and $f_i$ as the fraction of utterances routed to expert $i$.

The auxiliary loss is defined as:
\begin{equation}
\mathcal{L}_{aux} = E \sum_{i=1}^{E} p_i f_i.
\end{equation}

\noindent Let $\mathcal{L}_{\text{BCE}}$ denote the binary cross-entropy classification loss. The final training objective is defined as
\begin{equation}
\mathcal{L}_{\text{total}} = \mathcal{L}_{\text{BCE}} + \lambda \mathcal{L}_{aux}.
\end{equation}

\section{Experimental setup}
\label{sec:exp_setup}

This section describes the experimental protocol used to train and evaluate the proposed approach.

\subsection{Datasets and metrics}

\noindent \textbf{Training datasets:} We train our model using the 6 spoofing corpora 
ASVspoof5~\cite{wang2024asvspoof}, ASVspoof2019 LA~\cite{nautsch2021asvspoof}, ADD22~\cite{yi2022add}, FakeOrReal ~\cite{reimao2019dataset}, Codecfake~\cite{xie2025codecfake}, and MLADD~\cite{muller2024mlaad}. 
These datasets cover a wide range of spoofing conditions, synthesis methods, languages, and recording conditions, enabling the model to learn robust and generalizable representations. The distribution of spoof and bonafide samples in these corpora is reported below (Table~\ref{tab:corpora_distribution}).


\begin{table}[H]
\centering
\caption{Distribution of spoof and bonafide samples across the training corpora.}
\label{tab:train_corpora}
\resizebox{\linewidth}{!}{
\begin{tabular}{l
                S[table-format=6.0]
                S[table-format=6.0]
                S[table-format=7.0]}
\toprule
\textbf{Dataset} & {\textbf{Spoof}} & {\textbf{Bonafide}} & {\textbf{Total}} \\
\midrule
ASVspoof5       & 163560 &  18797 & 182357 \\
ASVspoof2019 LA &  22800 &   2580 &  25380 \\
ADD2022         &  24072 &   3012 &  27084 \\
FakeOrReal      &  26927 &  26939 &  53866 \\
Codecfake       & 634926 & 105821 & 740747 \\
MLAAD           & 243000 & 139085 & 382085 \\
\midrule
\textbf{Total}  & \multicolumn{1}{r}{\textbf{\num{1115285}}}
                & \multicolumn{1}{r}{\textbf{\num{296234}}}
                & \multicolumn{1}{r}{\textbf{\num{1411519}}} \\
\bottomrule
\end{tabular}
}
\label{tab:corpora_distribution}
\end{table}

\vspace{10pt}
\noindent \textbf{Development datasets:} Model checkpoint selection is performed using the development splits of several training corpora. Specifically, we use the official development sets of Codecfake~\cite{xie2025codecfake}, ASVspoof2019 LA~\cite{nautsch2021asvspoof}, ASVspoof5~\cite{wang2024asvspoof}, and ADD2022~\cite{yi2022add}.

\vspace{10pt} \noindent \textbf{Evaluation datasets:} For evaluation, we consider 14 corpora to assess generalization under a wide variability of voice generation and manipulation methods. ASVspoof2019 LA~\cite{nautsch2021asvspoof}, ASVspoof2021 LA, ASVspoof2021 DF~\cite{yamagishi2021asvspoof}, and ASVspoof5~\cite{wang2024asvspoof} include a large variety of TTS and VC techniques.  
Sonar~\cite{li2024sonar} and FakeOrReal~\cite{reimao2019dataset} focus on recent TTS systems, while DFADD~\cite{du2024dfadd} evaluates diffusion and flow-matching-based generation. 
Codecfake~\cite{xie2025codecfake} and LibriSeVoc~\cite{sun2023ai} target codec and vocoder-related artifacts respectively. 
ADD2022~\cite{yi2022add} (Track 1 and Track 3/Round 2) and ADD2023~\cite{yi2023add} (Track 1/Round 1 and Track 1/Round 2) provide Chinese-language spoofing conditions, and InTheWild~\cite{muller2022does} assesses performance in unconstrained real-world scenarios. Our evaluation follows the protocol introduced in Speech DF Arena~\cite{dowerah2026speech}. Accordingly, for the Sonar corpus we adopt the same preprocessing strategy and exclude samples generated with the SeedTTS synthesizer from the evaluation.

\vspace{10pt}
\noindent \textbf{Evaluation metrics:} We evaluate performance using the Equal Error Rate (EER). The EER corresponds to the operating point where the false acceptance rate is equal to the false rejection rate.
To assess cross-dataset generalization, we report both macro and micro EER across the test corpora. The micro (pooled) EER is computed by pooling the scores from all datasets and estimating a single global EER. The macro EER is computed as the average of the EER values obtained independently on each dataset, giving equal weight to all corpora, and is used as our primary metric.


\subsection{Implementation Details}

The toolkit used for the speaker embedding extractor is based on the Kiwano~\cite{rouvier2026} toolkit~\footnote{\url{https://github.com/kiwano-toolkit/kiwano}}. We train the model for 80k optimization steps using the AdamW optimizer. The initial learning rate is set to $2 \times 10^{-5}$ and decayed with a cosine annealing schedule to a final learning rate of $1 \times 10^{-6}$. A linear warmup is applied over the first 10\% of the total training steps to stabilize optimization in the early phase. Weight decay is set to $1 \times 10^{-2}$. Training is performed with a batch size of 128. Training audio samples are standardized to 2-second audio segments, obtained by randomly cropping a 2 s window from each training utterance at every iteration (when longer than 2 s).
The model is trained with a binary cross-entropy loss.
For the MoE configuration, the previously introduced auxiliary loss is weighted by $1 \times 10^{-2}$. At the beginning of training, the parameters of the self-supervised learning (SSL) backbone are frozen, and only the gating network, expert layers, and the MHFA module are optimized. The remaining SSL parameters are then progressively unfrozen following a linear schedule over the first 15\% steps.
To improve robustness, we apply on-the-fly data augmentation during training, including codec-based augmentation, additive noise augmentation using MUSAN, and reverberation augmentation via convolution with room impulse responses.

\section{Results}
\label{sec:results}


This section presents the experimental evaluation of the proposed MoE conversion. The used SSL backbone is first identified (Section~\ref{sec:ssl_select}), followed by a study of each architectural design choice (Sections~\ref{sec:moe_positions} to~\ref{sec:moe_topk}), and a comparison with LoRA-based alternatives (Section~\ref{sec:moe_lora}). Detailed per-corpus results are provided in Table~\ref{tab:detailed_scores}.

\subsection{SSL Backbone Selection}
\label{sec:ssl_select}


We evaluate three SSL backbone encoders: WavLM-Large~\cite{chen2022wavlm}, 
Wav2vec2-XLSR~\cite{baevski2020wav2vec} pre-trained on 128 languages, and 
HuBERT-Large~\cite{hsu2021hubert}. As shown in Table~\ref{tab:layer_wavlm}, 
WavLM-Large consistently achieves the lowest macro EER 
and is therefore selected as the backbone for all subsequent 
experiments.

Although WavLM-Large consists of 24 Transformer layers, 
we restrict MHFA to only use the first 13 layers, following 
the MHFA reference implementation~\cite{peng2023attention}. Since speech anti-spoofing relies predominantly on detecting low-level acoustic artifacts rather than linguistic information, restricting the model to the first 13 layers allows us to preserve the most relevant acoustic information. 

\begin{table}[H]
\centering
\caption{Baseline performance comparison using 13 or 24 layers (EER $\downarrow$ in \%).}
\label{tab:layer_wavlm}
\resizebox{\linewidth}{!}{%
\begin{tabular}{cccc}
\toprule
\textbf{SSL} & \textbf{Layer used} & \textbf{Macro EER (\%) $\downarrow$} & \textbf{Micro EER (\%) $\downarrow$} \\
\midrule
\multirow{2}{*}{WavLM Large} & 24 & 5.61 & 15.61 \\
  & 13 & \textbf{5.46} & 14.95 \\
\midrule
\multirow{2}{*}{Wav2vec2 XLSR}& 24 & \eer{6.014191754015575} & \eer{14.886249565102078} \\
 & 13 & \eer{6.008736345545781} & \eer{13.705235609355151} \\
\midrule
\multirow{2}{*}{HuBERT Large} & 24 & \eer{6.207722389159268} & \textbf{\eer{11.966452144478558}} \\
 & 13 & \eer{6.239249137666154} & \eer{12.90510995743307}\\
\bottomrule
\end{tabular}
}
\end{table}

As shown in Table~\ref{tab:layer_wavlm}, the 13-layer configuration achieves better performance for WavLM-Large, supporting this choice.

\begin{table*}[h]
  \caption{Detailed anti-spoofing results (EER $\downarrow$, in \%) with Codec = Codecfake and LSV = LibriSeVoc. The \textbf{MoE Config} columns specify the insertion position (first 6, last 6, all 13, or alternating), the number of experts $E$, the top-$k$ value, and the pooling strategy (att = attentive, stat = statistical). Macro EER averages EERs across corpora; Micro EER pools all trials. Best in \textbf{bold}, second-best \underline{underlined}.}
  \label{tab:detailed_scores}
  \centering
  \resizebox{\textwidth}{!}{%
  \begin{tabular}{l l c cccc cc cccc cc cc cccccc}
    \toprule
    \multirow{2}{*}{\textbf{Type}}
    & \multirow{2}{*}{\textbf{SSL}}
    & \multirow{2}{*}{\textbf{\#Params}}
    & \multicolumn{4}{c}{\textbf{MoE Config}}
    & \multicolumn{2}{c}{\textbf{Overall}}
    & \multicolumn{4}{c}{\textbf{ASVspoof}}
    & \multicolumn{2}{c}{\textbf{ADD22}}
    & \multicolumn{2}{c}{\textbf{ADD23}}
    & \multirow{2}{*}{\textbf{DFADD}} & \multirow{2}{*}{\textbf{Codec}} & \multirow{2}{*}{\textbf{FoR}} & \multirow{2}{*}{\textbf{LSV}} & \multirow{2}{*}{\textbf{ITW}} & \multirow{2}{*}{\textbf{Sonar}} \\
    \cmidrule(lr){4-7} \cmidrule(lr){8-9} \cmidrule(lr){10-13} \cmidrule(lr){14-15} \cmidrule(lr){16-17}
    & & & \textbf{Insertion} & \textbf{E} & \textbf{k} & \textbf{Pool}
    & \textbf{Macro} & \textbf{Micro}
    & \textbf{19LA} & \textbf{21DF} & \textbf{21LA} & \textbf{24}
    & \textbf{T1} & \textbf{T3-R2}
    & \textbf{T1-R1} & \textbf{T1-R2}
    & & & & & & \\
    \midrule
    
    \multirow{6}{*}{Baseline}
    & WavLM-L (13)   & \params{178204662} & --- & - & - & --- & \eer{5.46} & \eer{14.95} & \eer{0.04} & \underline{\eer{0.30}} & \eer{2.38} & \eer{17.12} & \eer{21.46} & \eer{5.50} & \eer{9.45} & \eer{16.34} & \textbf{\eer{0.00}} & \eer{2.05} & \textbf{\eer{0.00}} & \eer{0.21} & \eer{1.47} & \eer{0.14} \\
    & WavLM-L (24)   & \params{316769044} & --- & - & - & --- & \eer{5.61} & \eer{15.61} & \eer{0.04} & \eer{0.43} & \underline{\eer{2.32}} & \textbf{\eer{14.70}} & \eer{21.89} & \eer{6.04} & \eer{10.33} & \eer{18.61} & \textbf{\eer{0.00}} & \eer{2.09} & \textbf{\eer{0.00}} & \eer{0.39} & \eer{1.54} & \eer{0.21} \\
    & XLSR (13)      & \params{178196158} & --- & - & - & --- & \eer{6.01} & \eer{13.71} & \eer{0.33} & \eer{0.89} & \eer{3.60} & \eer{15.80} & \eer{22.70} & \eer{4.81} & \eer{8.67} & \eer{15.37} & \eer{0.13} & \eer{5.48} & \eer{0.35} & \eer{0.52} & \eer{1.84} & \eer{3.63} \\
    & XLSR (24)      & \params{316754644} & --- & - & - & --- & \eer{6.01} & \eer{14.89} & \eer{0.68} & \eer{1.31} & \eer{4.83} & \eer{17.89} & \eer{21.07} & \eer{5.26} & \eer{8.75} & \textbf{\eer{13.74}} & \eer{0.00} & \eer{5.35} & \eer{0.43} & \eer{0.98} & \eer{2.36} & \eer{1.53} \\
    & HuBERT-L (13)  & \params{178192574} & --- & - & - & --- & \eer{6.24} & \eer{12.91} & \eer{0.15} & \eer{0.72} & \eer{4.36} & \eer{16.62} & \eer{23.08} & \eer{4.85} & \eer{8.46} & \eer{15.38} & \textbf{\eer{0.00}} & \eer{1.57} & \textbf{\eer{0.00}} & \eer{3.10} & \eer{3.21} & \eer{5.85} \\
    & HuBERT-L (24)  & \params{316751060} & --- & - & - & --- & \eer{6.21} & \textbf{\eer{11.97}} & \eer{0.19} & \eer{1.02} & \eer{4.55} & \eer{16.23} & \eer{21.89} & \eer{4.45} & \eer{9.47} & \eer{17.57} & \eer{0.13} & \eer{1.92} & \eer{0.09} & \eer{2.30} & \eer{2.45} & \eer{4.66} \\
    \midrule
    
    \multirow{14}{*}{MoE} & \multirow{14}{*}{WavLM-L (13)}
      & \params{330128910} & first 6     & 4 & 1 & att  & \eer{5.60} & \eer{15.35} & \eer{0.05} & \eer{0.43} & \eer{3.14} & \underline{\eer{15.29}} & \eer{21.16} & \eer{6.13} & \eer{9.43} & \eer{17.01} & \eer{0.13} & \eer{2.26} & \textbf{\eer{0.00}} & \eer{0.88} & \eer{1.94} & \eer{0.56} \\
    &    & \params{330128910} & last 6      & 4 & 1 & att  & \eer{5.21} & \eer{13.80} & \eer{0.03} & \eer{0.30} & \eer{2.67} & \eer{17.34} & \eer{19.32} & \eer{4.88} & \eer{8.48} & \eer{15.46} & \textbf{\eer{0.00}} & \eer{2.44} & \textbf{\eer{0.00}} & \eer{0.38} & \eer{1.60} & \underline{\eer{0.08}} \\
    &   & \params{507373866} & all 13      & 4 & 1 & att  & \eer{5.77} & \eer{14.13} & \eer{0.12} & \eer{0.45} & \eer{3.09} & \eer{15.83} & \eer{20.85} & \eer{5.10} & \eer{9.86} & \eer{18.83} & \eer{0.13} & \eer{2.53} & \underline{\eer{0.04}} & \eer{0.87} & \eer{2.27} & \eer{0.76} \\
    &  & \params{355449618} & alternating & 4 & 1 & att  & \eer{5.42} & \eer{14.27} & \eer{0.04} & \eer{0.43} & \eer{2.62} & \eer{15.85} & \eer{20.42} & \eer{5.59} & \eer{9.33} & \eer{16.92} & \eer{0.10} & \eer{1.80} & \eer{0.13} & \eer{0.50} & \eer{1.86} & \eer{0.29} \\
    \cmidrule(lr){3-23}
    &    & \params{329340942} & last 6      & 4 & 1 & stat & \textbf{\eer{4.81}} & \underline{\eer{12.34}} & \eer{0.04} & \textbf{\eer{0.29}} & \eer{3.14} & \eer{16.28} & \textbf{\eer{18.01}} & \textbf{\eer{4.07}} & \textbf{\eer{7.89}} & \eer{14.46} & \eer{0.12} & \textbf{\eer{1.28}} & \underline{\eer{0.04}} & \eer{0.18} & \eer{1.49} & \underline{\eer{0.08}} \\
    &    & \params{329316366} & last 6      & 4 & 1 & max  & \underline{\eer{4.91}} & \eer{12.73} & \eer{0.05} & \eer{0.42} & \eer{3.20} & \eer{16.13} & \underline{\eer{18.70}} & \underline{\eer{4.26}} & \eer{8.18} & \eer{14.23} & \eer{0.13} & \eer{1.60} & \underline{\eer{0.04}} & \textbf{\eer{0.14}} & \eer{1.47} & \eer{0.14} \\
    &   & \params{329316366} & last 6      & 4 & 1 & mean & \eer{5.35} & \eer{13.75} & \eer{0.04} & \eer{0.43} & \eer{2.62} & \eer{15.81} & \eer{19.72} & \eer{5.23} & \eer{9.42} & \eer{17.63} & \textbf{\eer{0.00}} & \eer{1.82} & \underline{\eer{0.04}} & \eer{0.42} & \eer{1.62} & \underline{\eer{0.08}} \\
    \cmidrule(lr){3-23}
    &    & \params{227275716} & last 6      & 2 & 1 & stat & \eer{4.98} & \eer{12.81} & \eer{0.04} & \eer{0.30} & \eer{2.85} & \eer{16.09} & \eer{19.60} & \eer{4.65} & \eer{7.99} & \eer{14.96} & \textbf{\eer{0.00}} & \underline{\eer{1.30}} & \underline{\eer{0.04}} & \textbf{\eer{0.14}} & \eer{1.56} & \eer{0.14} \\
    &    & \params{277650378} & last 6      & 3 & 1 & stat & \eer{5.13} & \eer{13.81} & \eer{0.05} & \eer{0.43} & \eer{3.13} & \eer{16.84} & \eer{20.40} & \eer{4.94} & \underline{\eer{7.90}} & \underline{\eer{14.12}} & \eer{0.12} & \eer{1.90} & \eer{0.09} & \eer{0.17} & \eer{1.48} & \eer{0.19} \\
    &    & \params{277650378} & last 6      & 3 & 2 & stat & \eer{5.33} & \eer{13.07} & \eer{0.05} & \eer{0.43} & \eer{2.97} & \eer{16.56} & \eer{20.03} & \eer{5.32} & \eer{8.20} & \eer{16.54} & \eer{0.25} & \eer{2.50} & \textbf{\eer{0.00}} & \underline{\eer{0.16}} & \underline{\eer{1.39}} & \eer{0.27} \\
    &    & \params{329340942} & last 6      & 4 & 2 & stat & \eer{5.41} & \eer{13.99} & \eer{0.03} & \eer{0.43} & \textbf{\eer{2.28}} & \eer{17.01} & \eer{19.62} & \eer{5.66} & \eer{9.07} & \eer{17.66} & \underline{\eer{0.02}} & \eer{1.73} & \underline{\eer{0.04}} & \eer{0.40} & \eer{1.52} & \eer{0.27} \\
    &    & \params{329340942} & last 6      & 4 & 3 & stat & \eer{5.40} & \eer{14.46} & \textbf{\eer{0.03}} & \textbf{\eer{0.29}} & \eer{2.56} & \eer{17.23} & \eer{20.23} & \eer{5.52} & \eer{8.90} & \eer{16.02} & \textbf{\eer{0.00}} & \eer{2.75} & \textbf{\eer{0.00}} & \eer{0.45} & \eer{1.53} & \textbf{\eer{0.06}} \\
    &   & \params{378399702} & last 6      & 5 & 1 & stat & \eer{5.17} & \eer{13.60} & \underline{\eer{0.03}} & \eer{0.31} & \eer{2.85} & \eer{17.60} & \eer{19.30} & \eer{4.72} & \eer{8.43} & \eer{15.21} & \underline{\eer{0.02}} & \eer{1.97} & \textbf{\eer{0.00}} & \eer{0.32} & \eer{1.60} & \underline{\eer{0.08}} \\
    &    & \params{378399702} & last 6      & 5 & 2 & stat & \eer{5.00} & \eer{13.00} & \eer{0.04} & \eer{0.32} & \eer{2.78} & \eer{16.68} & \eer{19.56} & \eer{5.02} & \eer{7.96} & \eer{14.43} & \textbf{\eer{0.00}} & \eer{1.35} & \underline{\eer{0.04}} & \eer{0.17} & \textbf{\eer{1.36}} & \eer{0.35} \\
    &    & \params{378399702} & last 6      & 5 & 3 & stat & \eer{5.60} & \eer{14.23} & \eer{0.03} & \textbf{\eer{0.29}} & \eer{2.56} & \eer{16.67} & \eer{20.35} & \eer{5.67} & \eer{10.05} & \eer{18.22} & \eer{0.13} & \eer{1.95} & \underline{\eer{0.04}} & \eer{0.36} & \eer{1.61} & \eer{0.49} \\
    \bottomrule
  \end{tabular}
  }
\end{table*}

\subsection{Impact of MoE Layer Position}
\label{sec:moe_positions}

In the first experiment, we investigate the effect of inserting MoE layers at different positions within the 13 used Transformer layers of WavLM-Large. 
An attentive statistical pooling method is employed to produce the gating representation, and the routing strategy is fixed to top-$k=1$ (hard routing).
We evaluate four insertion strategies: early insertion (first six layers), late insertion (last six layers), full insertion (all 13 layers), and alternating insertion (one layer out of two).



\begin{table}[H]
\centering
\caption{Impact of MoE insertion position. All variants use $E=4$ experts, top-$k=1$ routing, and attentive pooling on the 13 used WavLM-Large layers. The \textbf{Insertion} column indicates which Transformer layers are converted to MoE layers.}
\label{tab:moe_positions}
\resizebox{\linewidth}{!}{
\begin{tabular}{l l c cc}
\toprule
\textbf{Type} & \textbf{SSL} & \textbf{Insertion} & \textbf{Macro EER (\%) $\downarrow$} & \textbf{Micro EER (\%) $\downarrow$} \\
\midrule
Baseline & WavLM-L (13) & ---         & \eer{5.462292486183278} & \eer{14.94920742998123} \\
\midrule
\multirow{4}{*}{MoE} & \multirow{4}{*}{WavLM-L (13)}
         & first 6     & \eer{5.601588853048004} & \eer{15.349807944474264} \\
         & & last 6      & \textbf{\eer{5.212828346460922}} & \textbf{\eer{13.8028395281045}} \\
         & & all 13      & \eer{5.765562921532959} & \underline{\eer{14.129139224491433}} \\
         & & alternating & \underline{\eer{5.4206937612215345}} & \eer{14.27229727897315} \\
\bottomrule
\end{tabular}
}
\end{table}

The best performance, as shown in Table~\ref{tab:moe_positions}, is consistently obtained when MoE layers are inserted in the last six layers. 
This suggests that such layers are more effective when applied to higher-level representations rather than low-level features. 

\subsection{Impact of the Pooling Strategy for Gating}
\label{sec:moe_pooling}

Using the configuration identified in the previous experiment where six MoE layers are inserted in the last Transformer layers, we now investigate how the pooling strategy used to compute the gating representation influences performance. The gating network requires an utterance-level embedding to determine which expert should be activated. The quality of this representation may therefore directly affect the routing decisions and, consequently, the overall system performance.

\noindent To study this aspect, we evaluate several pooling strategies, presented in Section~\ref{sec:gating_network}, that aggregate frame-level representations into a single utterance-level vector. In particular, we evaluate attentive statistical pooling, statistical pooling, max pooling and mean pooling.




\begin{table}[H]
\centering

\caption{Impact of the pooling strategy. All variants use $E=4$ experts and top-$k=1$ routing, with MoE layers inserted in the last 6 of the 13 used WavLM-Large layers.}
\label{tab:pooling_method}
\resizebox{\linewidth}{!}{
\begin{tabular}{lcccc}
\toprule
\textbf{Type} & \textbf{SSL} &  \textbf{Pooling} & \textbf{Macro EER (\%) $\downarrow$} & \textbf{Micro EER (\%) $\downarrow$} \\
\midrule
\multirow{4}{*}{MoE} 
            & \multirow{4}{*}{WavLM-L (13)} 
            & attentive & \eer{4.990163298}   & \eer{13.388614}    \\
            &           & statistical           & \textbf{\eer{4.812956148}}  & \textbf{\eer{12.33753383}}    \\
            &           & max                   & \underline{\eer{4.905493}} & \underline{\eer{12.731758}} \\
            &           & mean                  & \eer{5.348965672}   & \eer{13.75343557}    \\
\bottomrule
\end{tabular}
}
\end{table}

Table~\ref{tab:pooling_method} shows that simpler pooling strategies lead to improved performance compared to the attentive statistical pooling used in the previous configuration. In particular, both statistical pooling and max pooling yield lower EER values, with statistical pooling providing the best overall results. Based on these observations, we adopt statistical pooling as the default pooling strategy in the following experiments.

\subsection{Impact of the Number of Experts and Top-$k$}
\label{sec:moe_topk}

The last experiment was conducted by combining the best configurations from previous experiments. We fixed the 6 MoE layers as the last WavLM-Large layers and used statistical pooling as the gating pooling method.  
We analyze the effect of varying the number of experts $E$ and the top routing parameter $k$. These two parameters affect different aspects of the model. The routing parameter $k$ controls the inference cost by determining how many experts are activated per input, whereas the number of experts $E$ mainly impacts the model size.

\begin{table}[h]
\centering
\caption{Impact of the number of experts $E$ and top-$k$ routing. All variants use statistical pooling, with MoE layers inserted in the last 6 of the 13 used WavLM-Large layers.}
\label{tab:topk}
\resizebox{\linewidth}{!}{
\begin{tabular}{lcccccc}
\toprule
\textbf{Type} & \textbf{SSL} & \textbf{\#Params} & \textbf{E} & \textbf{k} & \textbf{Macro EER (\%) $\downarrow$} & \textbf{Micro EER (\%) $\downarrow$} \\
\midrule
\multirow{9}{*}{MoE} & \multirow{9}{*}{WavLM-L (13)} 
 & \params{227275716} & 2  & 1  & \underline{\eer{4.975247982601027}} & \underline{\eer{12.809104975334398}}\\
\cmidrule(lr){3-7}
 &  & \params{277650378} & 3  & 1 & \eer{5.126191190191797} & \eer{13.809814952743732} \\
 &  & \params{277650378} & 3  & 2 & \eer{5.334332625364797} & \eer{13.071184640710976} \\
\cmidrule(lr){3-7}
 &  &  \params{329340942} & 4  & 1 & \textbf{\eer{4.812956148}}  & \textbf{\eer{12.33753383}} \\
 &  &  \params{329340942} & 4  & 2 & \eer{5.4095759181701376} & \eer{13.99399171546778} \\
 & & \params{329340942} & 4  & 3  & \eer{5.397056335444265} & \eer{14.45753827453749}  \\
\cmidrule(lr){3-7}
 & &  \params{378399702} & 5  & 1  & \eer{5.17380424184079} & \eer{13.601911261540023}  \\
 & &  \params{378399702} & 5  & 2  & \eer{5.004603739088131} & \eer{13.001159744826266}  \\
 & &  \params{378399702} & 5  & 3  & \eer{5.602622648492562} & \eer{14.233962379875603}  \\
\bottomrule
\end{tabular}
}
\end{table}

Table~\ref{tab:topk} reports the impact of varying the number of experts $E$ and the routing parameter $k$. The best performance is obtained with $E=4$ experts and $k=1$, achieving a macro EER of 4.81\% and a micro EER of 12.34\%. Configurations with larger routing values ($k \geq 2$) tend to degrade performance, suggesting that activating multiple experts simultaneously may reduce the specialization effect expected from the MoE mechanism.
Interestingly, the configuration with $E=2$ experts and $k=1$ also achieves competitive performance while requiring a significantly smaller model (227M parameters). Although this setting does not reduce inference cost compared to $E=4$ (since $k=1$ in both cases), it provides a more memory-efficient alternative.

\subsection{Comparison with LoRA-based MoE Approaches}
\label{sec:moe_lora}
To validate our design choice using dense experts over LoRA-based experts adopted in prior works, we compare both strategies under identical conditions: MoE layers in the last six layers, statistical pooling, $E=4$ experts, and top-$k=1$ routing. In the LoRA-based setting, each layer expert is designed as a pair of LoRA modules applied to the two dense layers of the feed forward network block. Only these adapters and the gating network are trained, while the rest of the model remains frozen.

\begin{table}[H]
\centering
\caption{Comparison between full fine-tuning and LoRA-based adaptation for WavLM-Large MoE. Our MoE approach fine-tunes all 329M parameters, while LoRA-based approaches update only a small subset of a smaller model.}
\label{tab:lora_comparison}
\resizebox{\linewidth}{!}{
\begin{tabular}{l l cc c cc}
\toprule
\multirow{2}{*}{\textbf{Type}} & \multirow{2}{*}{\textbf{SSL}}
& \multicolumn{2}{c}{\textbf{\#Params}}
& \multirow{2}{*}{\textbf{Rank}}
& \multicolumn{2}{c}{\textbf{EER (\%)} $\downarrow$} \\
\cmidrule(lr){3-4} \cmidrule(lr){6-7}
& & \textbf{Total} & \textbf{Trainable} & & \textbf{Macro} & \textbf{Micro} \\
\midrule
MoE (ours) & WavLM-L (13) & \params{329340942} & \params{329340942} & -- & \textbf{\eer{4.812956148}} & \textbf{\eer{12.33753383}} \\
\midrule
\multirow{4}{*}{MoE-LoRA} & \multirow{4}{*}{WavLM-L (13)}
 & \params{178204662} & \params{2015256}  & 8  & \eer{6.803504920444066} & \eer{16.228352941004695} \\
 & & \params{182185998} & \params{3981336}  & 16 & \eer{6.841432425821956} & \eer{15.922949026292293} \\
 & & \params{186118158} & \params{7913496}  & 32 & \eer{6.72649215339650}  & \underline{\eer{15.786297885659595}} \\
 & & \params{193982478} & \params{15777816} & 64 & \underline{\eer{6.662098200360309}} & \eer{16.058918485502205} \\
\bottomrule
\end{tabular}
}
\end{table}

As shown in Table~\ref{tab:lora_comparison}, our MoE conversion,  at the cost of more parameters, outperforms LoRA-based approaches across all ranks, with LoRA performance improving as rank increases. This gap can be attributed to both the removal of the low-rank constraint on experts, and the fact that attention layers are jointly fine-tuned, allowing the model to co-adapt with the experts. While this comparison does not fully isolate the effect of the MoE structure from that of full fine-tuning, it suggests that both factors are important for expert expressiveness in this task.



\section{Analysis}
\label{sec:analysis}

\begin{figure}[h]
  \centering
  \resizebox{0.8\columnwidth}{!}{%
    \begin{minipage}{\columnwidth}
      \centering
      \begin{tikzpicture}[trim axis left]
\begin{groupplot}[
  group style={
    group size=1 by 6,
    vertical sep=0.8cm,
  },
  layerstyle,
]
 
\nextgroupplot[ylabel={$p(e{\mid}s,l{=}8)$}]
\addplot[fill=expA] coordinates {(0,0.14)(10,0.788136)(20,0.531250)(30,0.370000)(40,0.240000)(50,0.589474)(60,0.383333)};
\addplot[fill=expB] coordinates {(0,0.66)(10,0.042373)(20,0.093750)(30,0.250000)(40,0.12)(50,0.115789)(60,0.033333)};
\addplot[fill=expC] coordinates {(0,0.07)(10,0.135593)(20,0.187500)(30,0.015000)(40,0.28)(50,0.052632)(60,0.06)};
\addplot[fill=expD] coordinates {(0,0.13)(10,0.033898)(20,0.187500)(30,0.365000)(40,0.36)(50,0.242105)(60,0.523333)};
\draw[densely dotted,black!45,line width=0.7pt](axis cs:-5,0.25)--(axis cs:65,0.25);
\coordinate (top1) at (rel axis cs:0.5,1);
\coordinate (bot1) at (rel axis cs:0.5,0);
 
\nextgroupplot[ylabel={$p(e{\mid}s,l{=}9)$}]
\addplot[fill=expA] coordinates {(0,0.04)(10,0.067797)(20,0.125000)(30,0.046667)(40,0.00)(50,0.242105)(60,0.16)};
\addplot[fill=expB] coordinates {(0,0.08)(10,0.237288)(20,0.437500)(30,0.43)(40,0.48)(50,0.494737)(60,0.43)};
\addplot[fill=expC] coordinates {(0,0.11)(10,0.593220)(20,0.437500)(30,0.136667)(40,0.40)(50,0.210526)(60,0.398333)};
\addplot[fill=expD] coordinates {(0,0.77)(10,0.101695)(20,0.00)(30,0.386667)(40,0.12)(50,0.052632)(60,0.011667)};
\draw[densely dotted,black!45,line width=0.7pt](axis cs:-5,0.25)--(axis cs:65,0.25);
\coordinate (top2) at (rel axis cs:0.5,1);
\coordinate (bot2) at (rel axis cs:0.5,0);
 
\nextgroupplot[ylabel={$p(e{\mid}s,l{=}10)$}]
\addplot[fill=expA] coordinates {(0,0.02)(10,0.279661)(20,0.156250)(30,0.281667)(40,0.28)(50,0.305263)(60,0.36)};
\addplot[fill=expB] coordinates {(0,0.32)(10,0.279661)(20,0.031250)(30,0.30)(40,0.20)(50,0.178947)(60,0.165000)};
\addplot[fill=expC] coordinates {(0,0.11)(10,0.330508)(20,0.312500)(30,0.09)(40,0.20)(50,0.252632)(60,0.123333)};
\addplot[fill=expD] coordinates {(0,0.55)(10,0.110169)(20,0.500000)(30,0.328333)(40,0.32)(50,0.263158)(60,0.351667)};
\draw[densely dotted,black!45,line width=0.7pt](axis cs:-5,0.25)--(axis cs:65,0.25);
\coordinate (top3) at (rel axis cs:0.5,1);
\coordinate (bot3) at (rel axis cs:0.5,0);
 
\nextgroupplot[ylabel={$p(e{\mid}s,l{=}11)$}]
\addplot[fill=expA] coordinates {(0,0.28)(10,0.279661)(20,0.093750)(30,0.273333)(40,0.28)(50,0.20)(60,0.205000)};
\addplot[fill=expB] coordinates {(0,0.15)(10,0.025424)(20,0.062500)(30,0.330000)(40,0.20)(50,0.294737)(60,0.143333)};
\addplot[fill=expC] coordinates {(0,0.32)(10,0.610169)(20,0.750000)(30,0.32)(40,0.48)(50,0.421053)(60,0.471667)};
\addplot[fill=expD] coordinates {(0,0.25)(10,0.084746)(20,0.093750)(30,0.076667)(40,0.04)(50,0.084211)(60,0.18)};
\draw[densely dotted,black!45,line width=0.7pt](axis cs:-5,0.25)--(axis cs:65,0.25);
\coordinate (top4) at (rel axis cs:0.5,1);
\coordinate (bot4) at (rel axis cs:0.5,0);
 
\nextgroupplot[ylabel={$p(e{\mid}s,l{=}12)$}]
\addplot[fill=expA] coordinates {(0,0.71)(10,0.194915)(20,0.125000)(30,0.505000)(40,0.20)(50,0.178947)(60,0.035000)};
\addplot[fill=expB] coordinates {(0,0.20)(10,0.355932)(20,0.468750)(30,0.206667)(40,0.08)(50,0.473684)(60,0.611667)};
\addplot[fill=expC] coordinates {(0,0.05)(10,0.033898)(20,0.218750)(30,0.181667)(40,0.40)(50,0.136842)(60,0.155000)};
\addplot[fill=expD] coordinates {(0,0.04)(10,0.415254)(20,0.187500)(30,0.106667)(40,0.32)(50,0.210526)(60,0.198333)};
\draw[densely dotted,black!45,line width=0.7pt](axis cs:-5,0.25)--(axis cs:65,0.25);
\coordinate (top5) at (rel axis cs:0.5,1);
\coordinate (bot5) at (rel axis cs:0.5,0);
 
\nextgroupplot[ylabel={$p(e{\mid}s,l{=}13)$}]
\addplot[fill=expA] coordinates {(0,0.01)(10,0.067797)(20,0.218750)(30,0.753333)(40,0.08)(50,0.305263)(60,0.191667)};
\addplot[fill=expB] coordinates {(0,0.05)(10,0.271186)(20,0.281250)(30,0.086667)(40,0.24)(50,0.263158)(60,0.428333)};
\addplot[fill=expC] coordinates {(0,0.80)(10,0.135593)(20,0.281250)(30,0.033333)(40,0.48)(50,0.273684)(60,0.198333)};
\addplot[fill=expD] coordinates {(0,0.14)(10,0.525424)(20,0.218750)(30,0.126667)(40,0.20)(50,0.157895)(60,0.181667)};
\draw[densely dotted,black!45,line width=0.7pt](axis cs:-5,0.25)--(axis cs:65,0.25);
\coordinate (top6) at (rel axis cs:0.5,1);
\coordinate (bot6) at (rel axis cs:0.5,0);
 
\end{groupplot}
 
\node[font=\tiny\bfseries, anchor=south, yshift=-7pt] at (top1) {Layer 8};
\node[font=\tiny\bfseries, anchor=south, yshift=-7pt] at (top2) {Layer 9};
\node[font=\tiny\bfseries, anchor=south, yshift=-7pt] at (top3) {Layer 10};
\node[font=\tiny\bfseries, anchor=south, yshift=-7pt] at (top4) {Layer 11};
\node[font=\tiny\bfseries, anchor=south, yshift=-7pt] at (top5) {Layer 12};
\node[font=\tiny\bfseries, anchor=south, yshift=-7pt] at (top6) {Layer 13};
 
\node[anchor=north] at ($(current bounding box.south)+(0.6,0.1cm)$) {%
  \begin{tikzpicture}[font=\tiny]
    \fill[expA] (0,0)   rectangle (0.20,0.20); \node[right] at (0.25,0.10) {Expert 1};
    \fill[expB] (1.7,0) rectangle (1.90,0.20); \node[right] at (1.95,0.10) {Expert 2};
    \fill[expC] (3.4,0) rectangle (3.60,0.20); \node[right] at (3.65,0.10) {Expert 3};
    \fill[expD] (5.1,0) rectangle (5.30,0.20); \node[right] at (5.35,0.10) {Expert 4};
  \end{tikzpicture}%
};
 
\end{tikzpicture}
    \end{minipage}
  }
  \caption{Expert activation distribution $p(e \mid s, l)$ per layer $l$ 
and Sonar synthesizer $s$.}
  \label{fig:expert_activation}
\end{figure}
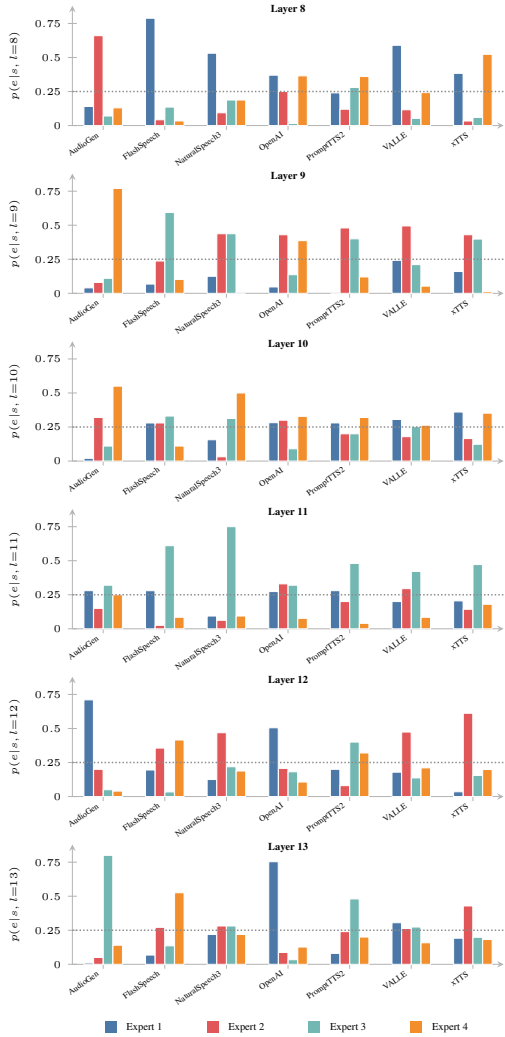

To assess whether MoE routing exhibits specialization with respect to particular synthesizers or, more broadly, attack types, we first analyze the distribution of expert activations for each synthesizer in the Sonar corpus. For each MoE layer \(l\), the activation counts of expert \(e\) for synthesizer \(s\) are normalized to obtain a probability distribution \(p(e|s,l)\).
Figure~\ref{fig:expert_activation} shows this distribution across the 6 MoE layers used in our best-performing model. 


\vspace{-1mm}

\vspace{-1mm}
The histograms reveal that expert activations remain relatively balanced across synthesizers. Although some synthesizers, such as AudioGen or OpenAI, show slightly different activation patterns, these differences do not indicate a clear routing specialization.


To quantitatively measure differences in expert routing across synthesis methods, we compute the Jensen–Shannon (JS) divergence between the expert activation distributions \(p(e|s,l)\) defined above. The JS divergence, which is based on the Kullback--Leibler divergence, quantifies the dissimilarity between two probability distributions: 

\vspace{-2mm}
\begin{equation}
JS(P \parallel Q) =
\frac{1}{2} KL(P \parallel M) +
\frac{1}{2} KL(Q \parallel M)
\end{equation}
where \(M = \frac{1}{2}(P + Q)\) and \(KL(\cdot)\) denotes the Kullback--Leibler divergence.
For each layer, pairwise JS divergence is computed for all synthesizer pairs, and the mean divergence is reported per dataset. This analysis is conducted on three test corpora: Sonar, ASVspoof2024 and Codecfake.

\begin{table}[h]
\scriptsize
\centering
\caption{Mean pairwise Jensen--Shannon divergence between synthesizer-specific expert activation distributions.}
\begin{tabular}{cccc}
\toprule
\textbf{MoE layer} & \textbf{Codecfake} & \textbf{ASV24} & \textbf{Sonar} \\
\midrule
8  & \js{0.085670} & \js{0.188726} & \js{0.273096} \\
\midrule
9  & \js{0.113055} & \js{0.170736} & \js{0.285966} \\
\midrule
10  & \js{0.120596} & \js{0.273338} & \js{0.210336} \\
\midrule
11  & \js{0.155543} & \js{0.210419} & \js{0.200574} \\
\midrule
12  & \js{0.158868}  & \js{0.252712} & \js{0.258125} \\
\midrule
13  & \js{0.161502} & \js{0.299017} & \js{0.291129} \\
\bottomrule
\label{tab:JS_mean}
\end{tabular}
\end{table}

Table~\ref{tab:JS_mean} summarizes the resulting JS mean divergences. 
For Codecfake and ASVspoof2024, we observe that the divergence slightly increases in deeper MoE layers. In contrast, Sonar shows a different trend, with divergence decreasing in the first layers before increasing again in the deepest ones. These trends suggest that routing patterns become slightly more distinct across synthesis methods in the deeper layers.
However, the magnitude of the mean divergence remains relatively low. This indicates that expert activation distributions remain largely similar across synthesizers and do not provide evidence of specialization to specific spoofing artifacts.  
More generally, interpreting the role of individual experts remains challenging. 
Even when variations in routing patterns are observed, it is difficult to associate them with specific attacks (spoofing artifacts), as experts may capture complex acoustic characteristics.


\section{Conclusion}
\label{sec:conclusion}

In this work, we proposed a method to convert a self-supervised speech model into a Mixture-of-Experts architecture for speech anti-spoofing. Unlike recent LoRA-based MoE approaches that keep the backbone frozen, our method replaces the dense feed-forward modules with full expert networks initialized from the pretrained weights, and jointly fine-tunes the entire model, prioritizing performance and robustness over parameter efficiency.
Validated on WavLM-Large across 14 evaluation corpora, the best configuration reduces the macro EER from 5.46\% to 4.81\%, an 11.9\% relative improvement over the baseline, and outperforms LoRA-based alternatives across all tested ranks.
Our analysis of expert activation patterns showed no clear specialization with respect to specific attacks. However, experts may capture complex acoustic patterns which are not straightforward to interpret.

Future work will conduct more in-depth analyses to better characterize potential expert specialization, and explore strategies to explicitly guide experts toward distinct spoofing methods. 

\newpage

\section{Acknowledgements}
This work was performed using HPC resources from GENCI–IDRIS (Grant 2025-AD011017195).

\bibliographystyle{IEEEtran}
\bibliography{Odyssey2026_BibEntries}

@inproceedings{rouvier2026, 
   title={{Kiwano: A Cutting-Edge Open-Source Toolkit for Speaker Verification}},
   booktitle={Odyssey 2026},
   author={Rouvier, Mickael and Bousquet, Pierre-Michel},
   year={2026},
}

@article{dowerah2026speech,
  title={Speech df arena: A leaderboard for speech deepfake detection models},
  author={Dowerah, Sandipana and Kulkarni, Atharva and Kulkarni, Ajinkya and Tran, Hoan My and Kalda, Joonas and Fedorchenko, Artem and Fauve, Benoit and Lolive, Damien and Alum{\"a}e, Tanel and Doss, Mathew Magimai-},
  journal={IEEE Open Journal of Signal Processing},
  year={2026},
  publisher={IEEE}
}

@article{vaccari2020deepfakes,
  title={Deepfakes and disinformation: Exploring the impact of synthetic political video on deception, uncertainty, and trust in news},
  author={Vaccari, Cristian and Chadwick, Andrew},
  journal={Social media+ society},
  volume={6},
  number={1},
  pages={2056305120903408},
  year={2020},
  publisher={SAGE Publications Sage UK: London, England}
}

@article{yi2023audio,
  title={Audio deepfake detection: A survey},
  author={Yi, Jiangyan and Wang, Chenglong and Tao, Jianhua and Zhang, Xiaohui and Zhang, Chu Yuan and Zhao, Yan},
  journal={arXiv preprint arXiv:2308.14970},
  year={2023}
}

@inproceedings{popov2021grad,
  title={Grad-tts: A diffusion probabilistic model for text-to-speech},
  author={Popov, Vadim and Vovk, Ivan and Gogoryan, Vladimir and Sadekova, Tasnima and Kudinov, Mikhail},
  booktitle={International conference on machine learning},
  pages={8599--8608},
  year={2021},
  organization={PMLR}
}

@inproceedings{mehta2024matcha,
  title={Matcha-TTS: A fast TTS architecture with conditional flow matching},
  author={Mehta, Shivam and Tu, Ruibo and Beskow, Jonas and Sz{\'e}kely, {\'E}va and Henter, Gustav Eje},
  booktitle={ICASSP 2024-2024 IEEE International Conference on Acoustics, Speech and Signal Processing (ICASSP)},
  pages={11341--11345},
  year={2024},
  organization={IEEE}
}

@article{wang2023neural,
  title={Neural codec language models are zero-shot text to speech synthesizers},
  author={Wang, Chengyi and Chen, Sanyuan and Wu, Yu and Zhang, Ziqiang and Zhou, Long and Liu, Shujie and Chen, Zhuo and Liu, Yanqing and Wang, Huaming and Li, Jinyu and others},
  journal={arXiv preprint arXiv:2301.02111},
  year={2023}
}

@inproceedings{tak2021end,
  title={End-to-end anti-spoofing with rawnet2},
  author={Tak, Hemlata and Patino, Jose and Todisco, Massimiliano and Nautsch, Andreas and Evans, Nicholas and Larcher, Anthony},
  booktitle={ICASSP 2021-2021 IEEE International Conference on Acoustics, Speech and Signal Processing (ICASSP)},
  pages={6369--6373},
  year={2021},
  organization={IEEE}
}

@inproceedings{jung2022aasist,
  title={Aasist: Audio anti-spoofing using integrated spectro-temporal graph attention networks},
  author={Jung, Jee-weon and Heo, Hee-Soo and Tak, Hemlata and Shim, Hye-jin and Chung, Joon Son and Lee, Bong-Jin and Yu, Ha-Jin and Evans, Nicholas},
  booktitle={ICASSP 2022-2022 IEEE international conference on acoustics, speech and signal processing (ICASSP)},
  pages={6367--6371},
  year={2022},
  organization={IEEE}
}

@article{ravanelli2018interpretable,
  title={Interpretable convolutional filters with sincnet},
  author={Ravanelli, Mirco and Bengio, Yoshua},
  journal={arXiv preprint arXiv:1811.09725},
  year={2018}
}

@article{tak2021end2,
  title={End-to-end spectro-temporal graph attention networks for speaker verification anti-spoofing and speech deepfake detection},
  author={Tak, Hemlata and Jung, Jee-weon and Patino, Jose and Kamble, Madhu and Todisco, Massimiliano and Evans, Nicholas},
  journal={arXiv preprint arXiv:2107.12710},
  year={2021}
}

@article{hsu2021hubert,
  title={Hubert: Self-supervised speech representation learning by masked prediction of hidden units},
  author={Hsu, Wei-Ning and Bolte, Benjamin and Tsai, Yao-Hung Hubert and Lakhotia, Kushal and Salakhutdinov, Ruslan and Mohamed, Abdelrahman},
  journal={IEEE/ACM transactions on audio, speech, and language processing},
  volume={29},
  pages={3451--3460},
  year={2021},
  publisher={IEEE}
}

@article{baevski2020wav2vec,
  title={wav2vec 2.0: A framework for self-supervised learning of speech representations},
  author={Baevski, Alexei and Zhou, Yuhao and Mohamed, Abdelrahman and Auli, Michael},
  journal={Advances in neural information processing systems},
  volume={33},
  pages={12449--12460},
  year={2020}
}

@article{chen2022wavlm,
  title={Wavlm: Large-scale self-supervised pre-training for full stack speech processing},
  author={Chen, Sanyuan and Wang, Chengyi and Chen, Zhengyang and Wu, Yu and Liu, Shujie and Chen, Zhuo and Li, Jinyu and Kanda, Naoyuki and Yoshioka, Takuya and Xiao, Xiong and others},
  journal={IEEE Journal of Selected Topics in Signal Processing},
  volume={16},
  number={6},
  pages={1505--1518},
  year={2022},
  publisher={IEEE}
}

@article{li2025measuring,
  title={Measuring the robustness of audio deepfake detectors},
  author={Li, Xiang and Chen, Pin-Yu and Wei, Wenqi},
  journal={arXiv preprint arXiv:2503.17577},
  year={2025}
}

@inproceedings{peng2023attention,
  title={An attention-based backend allowing efficient fine-tuning of transformer models for speaker verification},
  author={Peng, Junyi and Plchot, Old{\v{r}}ich and Stafylakis, Themos and Mo{\v{s}}ner, Ladislav and Burget, Luk{\'a}{\v{s}} and {\v{C}}ernock{\`y}, Jan},
  booktitle={2022 IEEE Spoken Language Technology Workshop (SLT)},
  pages={555--562},
  year={2023},
  organization={IEEE}
}

@article{mu2025comprehensive,
  title={A comprehensive survey of mixture-of-experts: Algorithms, theory, and applications},
  author={Mu, Siyuan and Lin, Sen},
  journal={arXiv preprint arXiv:2503.07137},
  year={2025}
}

@article{huang2024toward,
  title={Toward efficient inference for mixture of experts},
  author={Huang, Haiyang and Ardalani, Newsha and Sun, Anna and Ke, Liu and Lee, Hsien-Hsin S and Bhosale, Shruti and Wu, Carole-Jean and Lee, Benjamin},
  journal={Advances in Neural Information Processing Systems},
  volume={37},
  pages={84033--84059},
  year={2024}
}

@article{fedus2022switch,
  title={Switch transformers: Scaling to trillion parameter models with simple and efficient sparsity},
  author={Fedus, William and Zoph, Barret and Shazeer, Noam},
  journal={Journal of Machine Learning Research},
  volume={23},
  number={120},
  pages={1--39},
  year={2022}
}

@article{komatsuzaki2022sparse,
  title={Sparse upcycling: Training mixture-of-experts from dense checkpoints},
  author={Komatsuzaki, Aran and Puigcerver, Joan and Lee-Thorp, James and Ruiz, Carlos Riquelme and Mustafa, Basil and Ainslie, Joshua and Tay, Yi and Dehghani, Mostafa and Houlsby, Neil},
  journal={arXiv preprint arXiv:2212.05055},
  year={2022}
}

@inproceedings{fu2025ume,
  title={UME: Upcycling mixture-of-experts for scalable and efficient automatic speech recognition},
  author={Fu, Li and Yu, Shanyong and Li, Siqi and Fan, Lu and Wu, Youzheng and He, Xiaodong},
  booktitle={ICASSP 2025-2025 IEEE International Conference on Acoustics, Speech and Signal Processing (ICASSP)},
  pages={1--5},
  year={2025},
  organization={IEEE}
}

@article{hu2022lora,
  title={Lora: Low-rank adaptation of large language models.},
  author={Hu, Edward J and Shen, Yelong and Wallis, Phillip and Allen-Zhu, Zeyuan and Li, Yuanzhi and Wang, Shean and Wang, Liang and Chen, Weizhu and others},
  journal={Iclr},
  volume={1},
  number={2},
  pages={3},
  year={2022}
}

@article{pan2025molex,
  title={MoLEx: Mixture of LoRA Experts in Speech Self-Supervised Models for Audio Deepfake Detection},
  author={Pan, Zihan and Bhupendra, Sailor Hardik and Wu, Jinyang},
  journal={arXiv preprint arXiv:2509.09175},
  year={2025}
}

@inproceedings{laakkonen2025mixture,
  title={Mixture of Low-Rank Adapter Experts in Generalizable Audio Deepfake Detection},
  author={Laakkonen, Janne and Kukanov, Ivan and Hautam{\"a}ki, Ville},
  booktitle={2025 Asia Pacific Signal and Information Processing Association Annual Summit and Conference (APSIPA ASC)},
  pages={2211--2216},
  year={2025},
  organization={IEEE}
}

@article{chen2025adaptive,
  title={Adaptive Mixture of Low-Rank Experts for Robust Audio Spoofing Detection},
  author={Chen, Qixian and Xu, Yuxiong and Mandelli, Sara and Li, Sheng and Li, Bin},
  journal={IEEE Signal Processing Letters},
  year={2025},
  publisher={IEEE}
}

@inproceedings{hao2025wav2df,
  title={Wav2df-tsl: Two-stage learning with efficient pre-training and hierarchical experts fusion for robust audio deepfake detection},
  author={Hao, Yunqi and Chen, Yihao and Xu, Minqiang and Zhan, Jianbo and He, Liang and Fang, Lei and Fang, Sian and Liu, Lin},
  booktitle={2025 International Joint Conference on Neural Networks (IJCNN)},
  pages={1--8},
  year={2025},
  organization={IEEE}
}

@inproceedings{wang2025mixture,
  title={Mixture of experts fusion for fake audio detection using frozen wav2vec 2.0},
  author={Wang, Zhiyong and Fu, Ruibo and Wen, Zhengqi and Tao, Jianhua and Wang, Xiaopeng and Xie, Yuankun and Qi, Xin and Shi, Shuchen and Lu, Yi and Liu, Yukun and others},
  booktitle={ICASSP 2025-2025 IEEE International Conference on Acoustics, Speech and Signal Processing (ICASSP)},
  pages={1--5},
  year={2025},
  organization={IEEE}
}

@inproceedings{negroni2025leveraging,
  title={Leveraging mixture of experts for improved speech deepfake detection},
  author={Negroni, Viola and Salvi, Davide and Mezza, Alessandro Ilic and Bestagini, Paolo and Tubaro, Stefano},
  booktitle={ICASSP 2025-2025 IEEE International Conference on Acoustics, Speech and Signal Processing (ICASSP)},
  pages={1--5},
  year={2025},
  organization={IEEE}
}

@article{negroni2025attention,
  title={Attention-based Mixture of Experts for Robust Speech Deepfake Detection},
  author={Negroni, Viola and Salvi, Davide and Mezza, Alessandro Ilic and Bestagini, Paolo and Tubaro, Stefano},
  journal={arXiv preprint arXiv:2509.17585},
  year={2025}
}

@article{li2024sonar,
  title={Sonar: A synthetic ai-audio detection framework and benchmark},
  author={Li, Xiang and Chen, Pin-Yu and Wei, Wenqi},
  year={2024}
}

@inproceedings{reimao2019dataset,
  title={FoR: A Dataset for Synthetic Speech Detection.},
  author={Reimao, Ricardo and Tzerpos, Vassilios},
  booktitle={SpeD},
  pages={1--10},
  year={2019}
}

@inproceedings{yi2022add,
  title={Add 2022: the first audio deep synthesis detection challenge},
  author={Yi, Jiangyan and Fu, Ruibo and Tao, Jianhua and Nie, Shuai and Ma, Haoxin and Wang, Chenglong and Wang, Tao and Tian, Zhengkun and Bai, Ye and Fan, Cunhang and others},
  booktitle={ICASSP 2022-2022 IEEE International Conference on Acoustics, Speech and Signal Processing (ICASSP)},
  pages={9216--9220},
  year={2022},
  organization={IEEE}
}

@article{yi2023add,
  title={Add 2023: the second audio deepfake detection challenge},
  author={Yi, Jiangyan and Tao, Jianhua and Fu, Ruibo and Yan, Xinrui and Wang, Chenglong and Wang, Tao and Zhang, Chu Yuan and Zhang, Xiaohui and Zhao, Yan and Ren, Yong and others},
  journal={arXiv preprint arXiv:2305.13774},
  year={2023}
}

@article{xie2025codecfake,
  title={The codecfake dataset and countermeasures for the universally detection of deepfake audio},
  author={Xie, Yuankun and Lu, Yi and Fu, Ruibo and Wen, Zhengqi and Wang, Zhiyong and Tao, Jianhua and Qi, Xin and Wang, Xiaopeng and Liu, Yukun and Cheng, Haonan and others},
  journal={IEEE Transactions on Audio, Speech and Language Processing},
  volume={33},
  pages={386--400},
  year={2025},
  publisher={IEEE}
}

@inproceedings{muller2024mlaad,
  title={Mlaad: The multi-language audio anti-spoofing dataset},
  author={M{\"u}ller, Nicolas M and Kawa, Piotr and Choong, Wei Herng and Casanova, Edresson and G{\"o}lge, Eren and M{\"u}ller, Thorsten and Syga, Piotr and Sperl, Philip and B{\"o}ttinger, Konstantin},
  booktitle={2024 International Joint Conference on Neural Networks (IJCNN)},
  pages={1--7},
  year={2024},
  organization={IEEE}
}

@inproceedings{du2024dfadd,
  title={Dfadd: The diffusion and flow-matching based audio deepfake dataset},
  author={Du, Jiawei and Lin, I-Ming and Chiu, I-Hsiang and Chen, Xuanjun and Wu, Haibin and Ren, Wenze and Tsao, Yu and Lee, Hung-Yi and Jang, Jyh-Shing Roger},
  booktitle={2024 IEEE Spoken Language Technology Workshop (SLT)},
  pages={921--928},
  year={2024},
  organization={IEEE}
}

@article{nautsch2021asvspoof,
  title={ASVspoof 2019: Spoofing countermeasures for the detection of synthesized, converted and replayed speech},
  author={Nautsch, Andreas and Wang, Xin and Evans, Nicholas and Kinnunen, Tomi H and Vestman, Ville and Todisco, Massimiliano and Delgado, H{\'e}ctor and Sahidullah, Md and Yamagishi, Junichi and Lee, Kong Aik},
  journal={IEEE Transactions on Biometrics, Behavior, and Identity Science},
  volume={3},
  number={2},
  pages={252--265},
  year={2021},
  publisher={IEEE}
}

@article{yamagishi2021asvspoof,
  title={ASVspoof 2021: accelerating progress in spoofed and deepfake speech detection},
  author={Yamagishi, Junichi and Wang, Xin and Todisco, Massimiliano and Sahidullah, Md and Patino, Jose and Nautsch, Andreas and Liu, Xuechen and Lee, Kong Aik and Kinnunen, Tomi and Evans, Nicholas and others},
  journal={arXiv preprint arXiv:2109.00537},
  year={2021}
}

@article{wang2024asvspoof,
  title={ASVspoof 5: Crowdsourced speech data, deepfakes, and adversarial attacks at scale},
  author={Wang, Xin and Delgado, H{\'e}ctor and Tak, Hemlata and Jung, Jee-weon and Shim, Hye-jin and Todisco, Massimiliano and Kukanov, Ivan and Liu, Xuechen and Sahidullah, Md and Kinnunen, Tomi and others},
  journal={arXiv preprint arXiv:2408.08739},
  year={2024}
}

@article{muller2022does,
  title={Does audio deepfake detection generalize?},
  author={M{\"u}ller, Nicolas M and Czempin, Pavel and Dieckmann, Franziska and Froghyar, Adam and B{\"o}ttinger, Konstantin},
  journal={arXiv preprint arXiv:2203.16263},
  year={2022}
}

@inproceedings{sun2023ai,
  title={Ai-synthesized voice detection using neural vocoder artifacts},
  author={Sun, Chengzhe and Jia, Shan and Hou, Shuwei and Lyu, Siwei},
  booktitle={Proceedings of the IEEE/CVF Conference on Computer Vision and Pattern Recognition},
  pages={904--912},
  year={2023}
}

\end{document}